\documentclass[aps,pra,twocolumn,groupedaddress]{revtex4-1}

\pdfoutput=1

\usepackage{graphics}
\usepackage{epsfig}
\usepackage{color}
\usepackage{amsmath}
\usepackage{bm}

\begin{document}




\title{Analytical approximations for the higher energy structure in strong field ionization with inhomogeneous electric fields}

\author{L. Ortmann$^{1}$}
\email[]{ortmann@pks.mpg.de}
\author{A. S. Landsman$^{1,2}$}
\email[]{landsman@pks.mpg.de}
\affiliation{$^1$Max Planck Institute for the Physics of Complex Systems, N\"othnitzer Stra{\ss}e 38, D-01187 Dresden, Germany}
\affiliation{$^2$Department of Physics, Max Planck Postech, Pohang, Gyeongbuk 37673, Republic of Korea}

\date{\today}

\begin{abstract}
Recently, the emergence of a higher energy structure (HES) due to a spatial inhomogeneity in the laser electric field, as is typically found close to a nano tip, was reported in Phys.~Rev. Letter {\bf 119}, 053204 (2017).  For practical applications, such as the characterization of near-fields or the creation of localized sources of monoenergetic electron beams with tunable energies, further insight into the nature of this higher energy structure is needed. Here, we give a closed form analytical approximation to describe the movement of the electron in the inhomogeneous electric field. In particular, we derive a simple scaling law for the location of the HES peak and give a scheme to analytically tune the width of the peak, both of which will prove useful in optimizing the nanostructure size or geometry for creating the HES in experimental settings.

\end{abstract}

\maketitle

\section{Introduction}
Applying electric fields strong enough to compete with the Coulomb potential of atoms and molecules will lead to the liberation of electrons that were initially bound in such a potential. This process of ionization can be studied on the attosecond time-scale when the laser is chosen to be a pulse with a duration on a similarly short time scale \cite{krausz_09}. Experimental and theoretical studies of the electron's motion after ionization have led to the observation of many different phenomena, such as above threshold ionization, harmonic generation or double ionization \cite{agostini1979,corkum, lHuilier1993high, lewenstein94A, lHuilier1982multiply, staudte2007binary}, all of which helped reveal structural and dynamical information about the atomic or molecular system. 

One special way to produce short and highly intense electric fields is to focus a short laser pulse on a nanostructure, in the vicinity of which there will be a strong field enhancement of the incoming laser field due to plasmonic effects. The atomic or molecular gas to be ionized is then placed in this region of field enhancement \cite{kimNature, husakou2011theory, sivis, shaaran2013high, shaaran2012quantum, hiroelectron, ciappina2017attosecond, yavuz2012generation, he2013wavelength, ciappina2012above,choi}.  As the enhanced field decays exponentially with increasing spatial distance from the apex of the nanotip, it is vital to account for the strong spatial inhomogeneity when describing the movement of the electron after ionization. This stands in contrast to strong field effects observed in fields without plasmonic enhancement, where in most cases the dipole approximation can be applied and the laser field can be assumed to be spatially homogeneous. Consequently, in ionization close to nanostructures new effects are observed due to the spatial inhomogeneity and theoretical descriptions need to be adapted to include this local dependence in the laser electric field. 

\begin{figure}
\resizebox{3in}{!}{\includegraphics{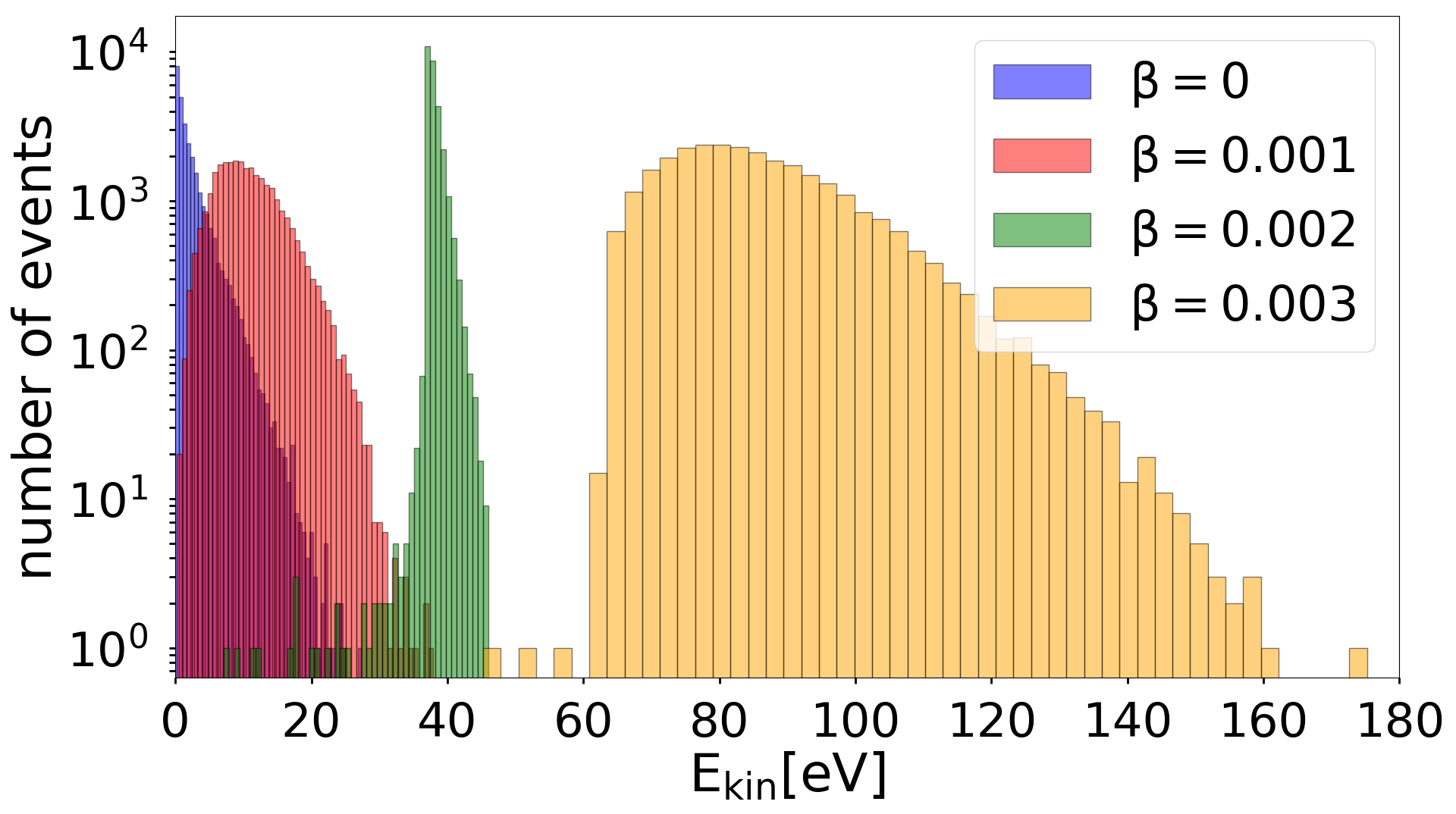}}  
\caption{Histograms of the final electron energy shows the accumulation of electrons at higher energies in an inhomogeneous field, where the inhomogeneity is quantified by the parameter $\beta$ given in the legend (see eq. \ref{eq:Efield} and the corresponding explanations in the text for details of the definition of $\beta$).  
The parameters are chosen as in \cite{ortmann2017emergence} with an intensity of $I=1 \cdot 10^{14} \mathrm{W/cm^2}$, the wavelength being $\lambda = 2 \mu m$ in a 2-cycle pulse ($N=2$) with a CEP of $\phi_0=\pi$ in Helium ($I_p =0.9 a.u.$).
}
\label{fig:HES_peak_for_different_beta}
\end{figure}

Recently, Ortmann and co-workers \cite{ortmann2017emergence} identified a higher energy structure (HES) arising due to spatial inhomogeneities in enhanced fields. In particular, electrons were found to accumulate at a relatively high energy, in the regime of $2 U_p$ with $U_p = I/(4 \omega)$ being the ponderomotive potential, forming a prominent hump in the spectrum. In Fig. \ref{fig:HES_peak_for_different_beta} this effect is displayed for four different degrees of inhomogeneity. As the energy at which this peak is located turns out to be very sensitive to the inhomogeneity of the field, this allows for tuning the energy around which the electrons accumulate by varying the geometry of the nanostructure.   
In addition, it is important for certain applications to have a narrow energy distribution in the HES peak in order to generate ultrashort and nearly monoenergetic electron beams. So at which inhomogeneity does the HES peak appear at minimal width?  Both issues, the position and width of the HES peak, were investigated with numerical simulations in
 \cite{ortmann2017emergence}.  However, a deeper analytic understanding for the dependence of the HES on field inhomogeneity is still missing, but is vital in order to tailor the parameters for potential applications.  The presence of inhomogeneous electric fields considerably complicates the equations of motion, making tractable analytic approximations difficult.  Hence, as of yet, to the best of our knowledge, there are no closed form analytical approximations describing electron motion in the vicinity of a nanotip.  
 
 

The rest of the work is organized as follows: after introducing the relevant equations of motion in inhomogeneous laser field in section \ref{sec:Inhom_field_description}, we derive an approximate analytic expression for electron acceleration as a function of time in 
section \ref{sec:General_analytical_solutions}.  Finding that this approximation matches the full solution very well, we also present a scheme to calculate the final velocity in an analytical framework. Then, in section \ref{sec:beta_dependence_of_final_energy}, we apply this knowledge to derive a simple scaling law showing how electron's final energy depends on field inhomogeneity.  Moreover, we find that there is an inhomogeneity parameter which leads to a minimal spectral width of the HES hump. We find that this minimum can be understood as due to cancellation of two competing effects:  electrons ionized earlier will be initially accelerated (relative to other electrons), but will experience greater deceleration later on in the pulse, depending on the degree of field inhomogeneity.  We also give a scheme for quickly estimating analytically this optimal inhomogeneity in section \ref{sec:beta_dependence_of_HES_peak_width}. Finally, section \ref{sec:Conclusions} concludes and summarizes.  

\section{Theoretical description of the inhomogeneous field} \label{sec:Inhom_field_description}
The HES observed in \cite{ortmann2017emergence} was in the intermediate regime $\delta = l_F/l_q \sim 1$ (where $l_F$ is the field decay length and $l_q$ is the electron quiver amplitude) \cite{herink2012}, whereby the quiver amplitude is large enough to experience the field inhomogeneity, but small enough to remain in the vicinity of the nanotip during the duration of the laser pulse.  
As the aim of the present work is to gain a deeper understanding of the HES arising under these conditions, we will stay in the same regime.  Note that this intermediate regime is also considerably more analytically challenging, since the electric field can neither be assumed to be spatially homogeneous (as is the case for small electron quiver amplitude, $\delta \gg 1$) nor static in the vicinity of a nanotip (as is the case when the electron quickly escapes the nanostructure, $\delta \ll 1$) \cite{herink2012}.
For concreteness, we use a laser intensity of $I=1E14$ and a wavelength of $\lambda = 2 \mu m$. The exponentially decaying inhomogeneous field is characterized by the decay length $l_F$, which we choose to be $8.3$ nm. Note that the decay length is, in the case of a nanosphere, in the order of its radius and thus gives an estimate of the size of the nanostructure needed to create such inhomogeneities \cite{ciappina2017attosecond}.

\begin{figure}
\resizebox{3in}{!}{\includegraphics{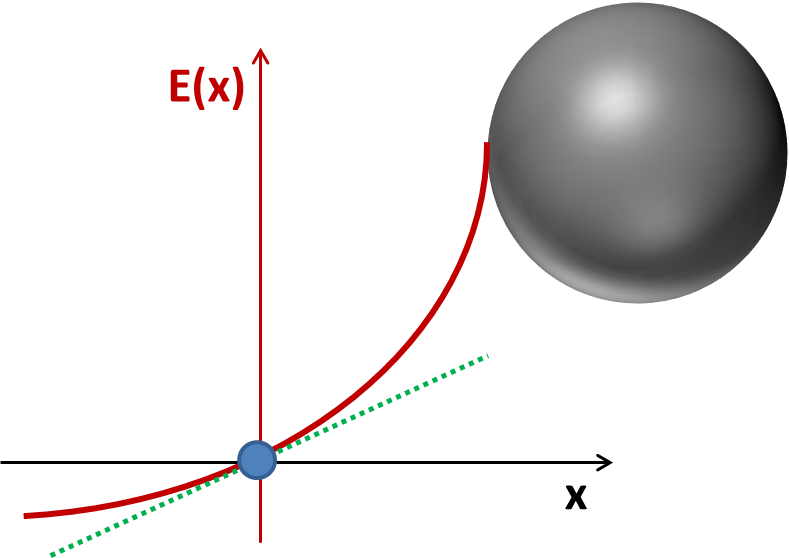}}  
\caption{Schematic picture of the linear approximation (green line) of the exponentially decaying field (red line) close to a nanostructure (grey sphere) according to eq. \ref{eq:Efield}. The linear approximation matches the exponential curve at the position of the atom to be ionized (blue circle), which is located at $x=0$.}
\label{fig:Inhom_field_sketch}
\end{figure}

For the description of the inhomogeneous laser field we apply a linear approximation, which is motivated and explained in more detail in \cite{ciappina2012above} and reads
\begin{equation}
{\bf E}(x,t)=E_0 (1+2 \beta x) f(t) \cos(\omega t+ \phi_0) \hat{x}
\label{eq:Efield}
\end{equation}
where  $E_0$ denotes the amplitude of the electric field at the location of the parent atom, which is placed at $x=0$, $\phi_0$ is the carrier-envelope phase (CEP) and $f(t)=\cos^{2}\left(\frac{\omega t}{2N }\right)$ describes the pulse envelope, where $N$ is a number of cycles in the pulse. 
The parameter $\beta$ characterizes the inhomogeneity and is given by the inverse $(1/e)$ decay length:  $2 \beta = 1/l_F$. For $\beta=0$ the case of a homogeneous field is restored. $\hat{x}$ is the direction of laser polarization with decreasing $x$ signifying a movement away from the tip. For a more intuitive understanding of this description of the field a sketch is provided in Fig. \ref{fig:Inhom_field_sketch}. Note that the linear approximation given in eq. \ref{eq:Efield} will fail if the term $(1+2 \beta x)$ changes sign due to sufficiently large negative values of $x$.

\section{General analytical solutions} \label{sec:General_analytical_solutions}

\subsection{Analytical approximation for the acceleration}
The comparison between TDSE and Classical trajectory Monte Carlo simulations (CTMC) in \cite{ortmann2017emergence} revealed that the sizeable gain in electron energy can be explained classically by the movement of the electron in the inhomogeneous field. Consequently, it is sufficient to solve Newton's equation of motion
\begin{equation}
 \ddot{{\bf{r}}}(t) = - {\bf E}(x,t) - \frac{{\bf r}}{r^3}.  \label{eq:ODE0}
\end{equation}
If we neglect the Coulomb force, which can be treated as a perturbative correction \cite{corkum}, the motion along the different coordinates decouples, with non-trivial motion being driven by the laser electric field along the x-direction:
\begin{equation}
  \ddot{x}(t) = -E_0 \cos^2\left(\frac{\omega t}{2 N}\right) \cos(\omega t +\phi_0) \cdot (1 + 2 \beta x). \label{eq:ODE1}
 \end{equation}
We now shift the x-coordinate such that the position dependent potential is centered around zero
\begin{equation}
  \ddot{x}_M(t) = -E_0 f(t)  \cos(\omega t + \phi_0) \cdot (2 \beta x_M), \label{eq:ODE2}
 \end{equation}
with the underlying transformation given by $x_{M} = x + 1/2 \beta$. 

For a slowly varying envelope relative to the laser frequency, or $1/2N \ll 1$, we can neglect the time-dependence in the envelope, $f(t)$, when solving the equation of motion.  
In this case, the solution is given by Mathieu functions \cite{mathieu1868memoire}. Using initial conditions of $x_M(t_0) = -I_p/E(x=0,t_0) + 1/2\beta$ and $\dot{x}_M(t_0) = 0$ at the time of ionization $t_0$ (with $t_0=0$ corresponding to ionization at the center of the pulse) the solution to equation \ref{eq:ODE2} reads

\begin{equation}
  x_{M}(t) \approx \frac{\left(  M_C'(\eta_0) M_S(\eta) -  M_C(\eta) M_S'(\eta_0) \right) A }{2\beta E_0 \left( M_C'(\eta_0) M_S(\eta_0) -M_C(\eta_0) M_S'(\eta_0) \right)}  \label{eq:x_Mathieu}
 \end{equation}
 where $\eta_0 = (0,\alpha,\gamma_0)$ and $\eta= (0,\alpha,\gamma)$ denote the three arguments of the even and odd Mathieu functions, $M_C$ and $M_S$ (respectively), as they are defined in Mathematica \cite{mathematica} and where $\alpha = - \frac{4 \beta E_0}{\omega^2}$, $\gamma = 1/2 (\phi_0 + \omega t)$, $\gamma_0 = 1/2 (\phi_0 + \omega t_0)$. $A$ is defined as $A = \left(E_0 - 2 \beta I_p \sec^2{(\frac{\omega t_0}{2 N})} \sec(\phi_0 + \omega t_0) \right)$. Moreover, the prime denotes the derivative of the Mathieu function with respect to time, which can, just as the Mathieu functions themselves, be evaluated with the help of standard libraries.  
 
\begin{figure}
\resizebox{3in}{!}{\includegraphics[angle=0]{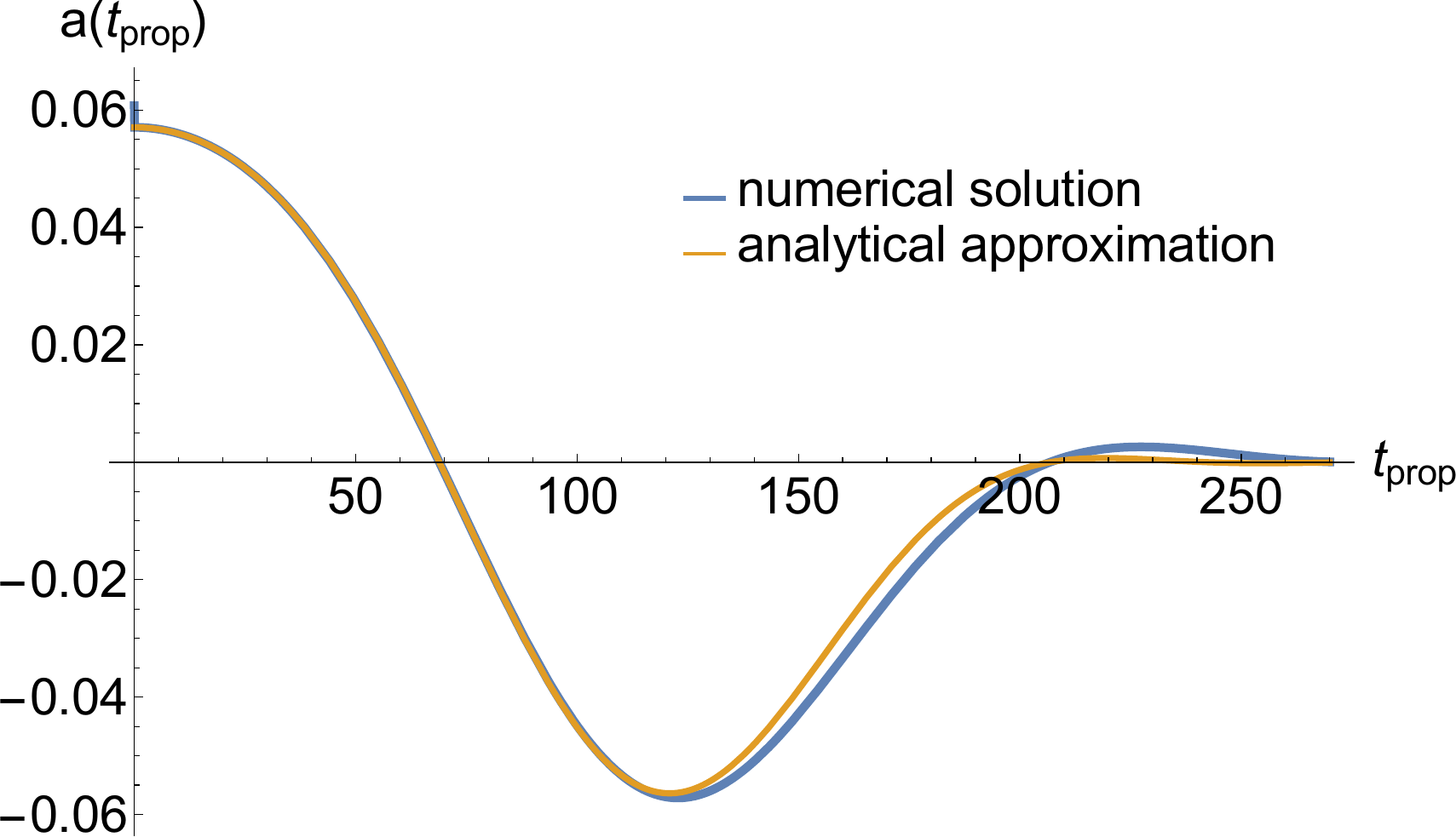}}  
\caption{Full numerical solution of eq.  \ref{eq:ODE2} for the acceleration $a$ (blue) compared to the approximate analytical solution (orange) according to eq. \ref{eq:x_Mathieu} plugged into eq. \ref{eq:ODE2} as a function of the propagation time $t_{prop}$ after ionization, which was chosen to occur at $t_0 = 0$ for this trajectory. The parameters are chosen as in \cite{ortmann2017emergence} with an intensity of $I=1 \cdot 10^{14} \mathrm{W/cm^2}$, the wavelength being $\lambda = 2 \mu m$ in a 2-cycle pulse ($N=2$) with a CEP of $\phi_0=\pi$ in Helium ($I_p =0.9$ a.u.) and the inhomogeneity is given by $\beta=0.002$.}
\label{fig:acceleration_comparison}
\end{figure}

Fig. \ref{fig:acceleration_comparison} compares the acceleration obtained from the full numerical solution of eq. \ref{eq:ODE1} with the above analytic approximation for the position plugged into the RHS of eq. \ref{eq:ODE2}. The agreement between the two is surprisingly good, given that only a two cycle ($N=2$) pulse is used.  An even better agreement should be expected for longer pulses, where the slowly-varying envelope assumption is more valid.  

Also note that eq.  \ref{eq:ODE2}, with $x_{M}(t)$ given by eq. \ref{eq:x_Mathieu}, serves as a closed form approximate analytic expression for the time-dependent force experienced by the electron following ionization.  Up till now, due to the spatial field inhomogeneity, this force could only be calculated by numerically solving the equation of motion (eq. \ref{eq:ODE1}) first.


\subsection{Analytical approximation for the velocity}
We now focus on the analytical description of electron's velocity following ionization. This is of particular interest, in this context, as the velocity at the end of the pulse determines the final energy of electrons forming the HES. As Fig. \ref{fig:HES_peak_for_different_beta} already reveals, both the peak position and the width depend strongly on the inhomogeneity parameter $\beta$, but so far we lack a quantitative description of this dependence. In order to be able to derive such dependence, we first need to understand how the electron's velocity evolves in the inhomogeneous field in general and how we can describe it analytically.

A first approximation of the velocity is to merely derive the approximation of the position according to eq. \ref{eq:x_Mathieu} with respect to time, which yields
\begin{equation}
 v_M(t)= \omega \frac{\left( M_C'(\eta_0) M_S'(\eta) - M_C'(\eta) M_S'(\eta_0) \right) A }{ 2\beta E_0 \left( M_C'(\eta_0) M_S(\eta_0) -M_C(\eta_0) M_S'(\eta_0) \right) } \label{eq:v_Mathieu}
\end{equation}
following the same notation as in eq. \ref{eq:x_Mathieu}.
As this result poorly matches the numerical solution for the velocity, further refinement is needed. To this end, we approximate the velocity by integrating the acceleration we obtain when plugging eq. \ref{eq:x_Mathieu} into eq. \ref{eq:ODE2}, using integration by parts and dropping the terms that contain derivatives of the envelope $f(t)$, assuming a slowly-varying envelope (see Appendix \ref{sec:appendix_integration_by_parts} for details):
\begin{equation}
 v_{approx1}(t) =  \left[ v_M(t') \cdot f(t') - x_M(t') \cdot  f(t') \cdot \frac{d}{dt'} f(t')   \right]_{t_0}^{t}. \label{eq:vApprox1}
\end{equation}

\begin{figure}
\resizebox{3in}{!}{\includegraphics[angle=0]{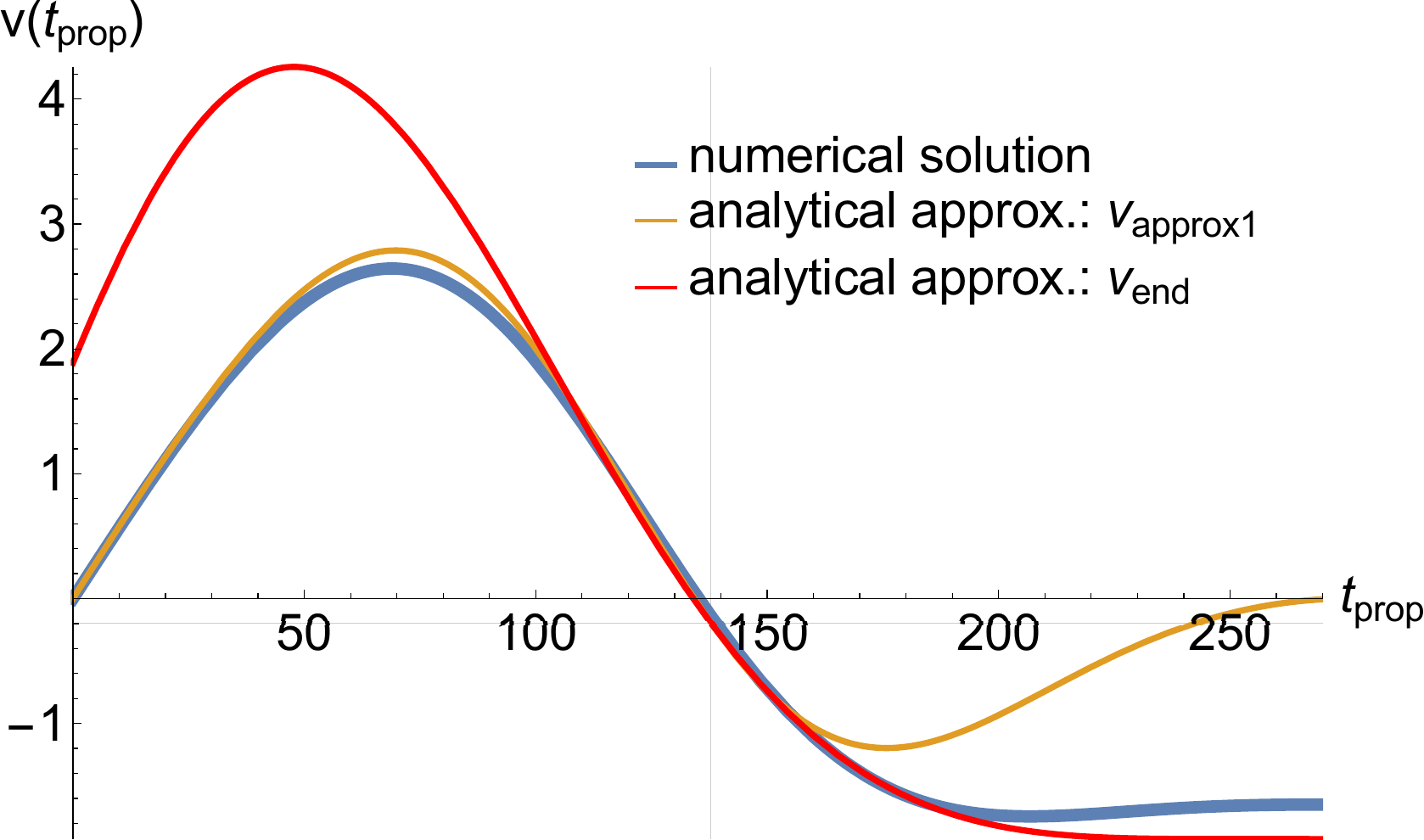}}  
\caption{Numerical solution for the velocity $v$ (blue, solution of eq. \ref{eq:ODE1}) compared to the approximate analytical solution for the first part of propagation (orange line, eq. \ref{eq:vApprox1}) and for the second part of propagation (red line, eq. \ref{eq:v_end}) as a function of the propagation time $t_{prop}$ after ionization, which was chosen to occur at $t_0 = 0$ for this trajectory. All parameters were chosen to be the same as those in Fig. \ref{fig:acceleration_comparison}.
The initial conditions for eq. \ref{eq:v_end} were chosen at time $t_1 = \pi/\omega$, one half-cycle away from the center of the pulse and $\omega_1$ was chosen as $\omega/(2 N)\cdot 1.1$. The gray vertical and horizontal line mark the parameters $t_1$ and $v_1$, respectively.
}
\label{fig:velocity_Approx1_vEnd}
\end{figure}

The result of this approximation is depicted as orange line in Fig. \ref{fig:velocity_Approx1_vEnd}. We find that this approximation matches the numerical solution of eq. \ref{eq:ODE1} (blue line) nicely in the first half of propagation, but deviates considerably in the second half.  The reason for this is the break-down of the slowly-varying envelope assumption for short-cycle pulses.  In particular, this assumption works well near the center of the pulse envelope where the time derivative is small, but breaks down for sufficiently short pulses later on, as the derivative of $f(t)$ becomes larger.  (Note that most electron trajectories come from ionization near the peak of the laser pulse envelope, where the derivative of $f(t)$ is zero.)  



To take account of this, we include the time-dependence of the pulse envelope at later times in electron propagation, but drop the primary oscillation of the laser field, in order to keep the equation analytically tractable:  

\begin{equation}
 \ddot{x}_{end}(t) = -E_1 \cdot (2 \beta x_{end}(t)) \cdot  \cos^2(\omega_1 t) \label{eq:ODE_env}
\end{equation}
with initial conditions $v_1 = \dot{x}_{end}(t_1) = v_{approx1}(t_1)$ and $x_1 = x_{end}(t_1) = x_M (t_1) - 1/(2 \beta)$, where $t_1$ is chosen as the time where the first part of the solution $v_{approx1}$ starts to deviate considerably from the numerical solution of eq. \ref{eq:ODE1}. $E_1$ is chosen such that the solution $v_{approx1}$ and $v_{end}$ match at $t_1$ and consequently is given by $E_1 = - a_{M}(t_1)f(t_1)/(2 \beta x_{end}(t_1) \cos^2(\omega_1 t_1)) $ with $a_M$ denoting the second time derivative of $x_M$.
The solution $x_{end}$ of eq. \ref{eq:ODE_env} is again obtained by employing the Mathieu functions and this solution for the electron's position can be easily derived with respect to time, $v_{end} = d/dt ~ x_{end}$, thus obtaining the following approximation for the velocity of the electron at the end of the pulse:
\begin{widetext}
\begin{equation}
 v_{end}(t) = \frac{\left(-v1 M_C(\zeta_0) + \omega_1 x_1 M_C'(\zeta_0) \right) \cdot M_S'(\zeta) + M_C(\zeta) \cdot \left( v_1 M_S(\zeta_0) - \omega_1 x_1 M_S'(\zeta_0) \right)}{ M_c'(\zeta_0) M_S(\zeta_0) - M_C(\zeta_0) M_S'(\zeta_0) }, \label{eq:v_end}
\end{equation}
\end{widetext}
where $\zeta$ denotes the following three arguments of the respective Mathieu function $(\frac{\beta E_1}{\omega_1^2}, - \frac{\beta E_1}{2 \omega_1^2}, \omega_1 t)$ and analogously  $\zeta_0$ replaces $(\frac{\beta E_1}{\omega_1^2}, - \frac{\beta E_1}{2 \omega_1^2}, \omega_1 t_1)$.
As Fig. \ref{fig:velocity_Approx1_vEnd} shows, this approximate solution (red line) matches well the second part of the electron's propagation in the laser field and, in particular, it also allows for calculating the final kinetic energy at the end the pulse. 

Hence we find that an analytic approximation given by eq. \ref{eq:v_end}, where the explicit time-dependence of the laser field is approximated by the pulse envelope, gives an accurate estimate for the final electron energy (even though it may deviate considerably from the exact solution during the propagation).

\section{Dependence of the final energy on the inhomogeneity parameter} \label{sec:beta_dependence_of_final_energy}
\begin{figure}
\resizebox{3in}{!}{\includegraphics[angle=0]{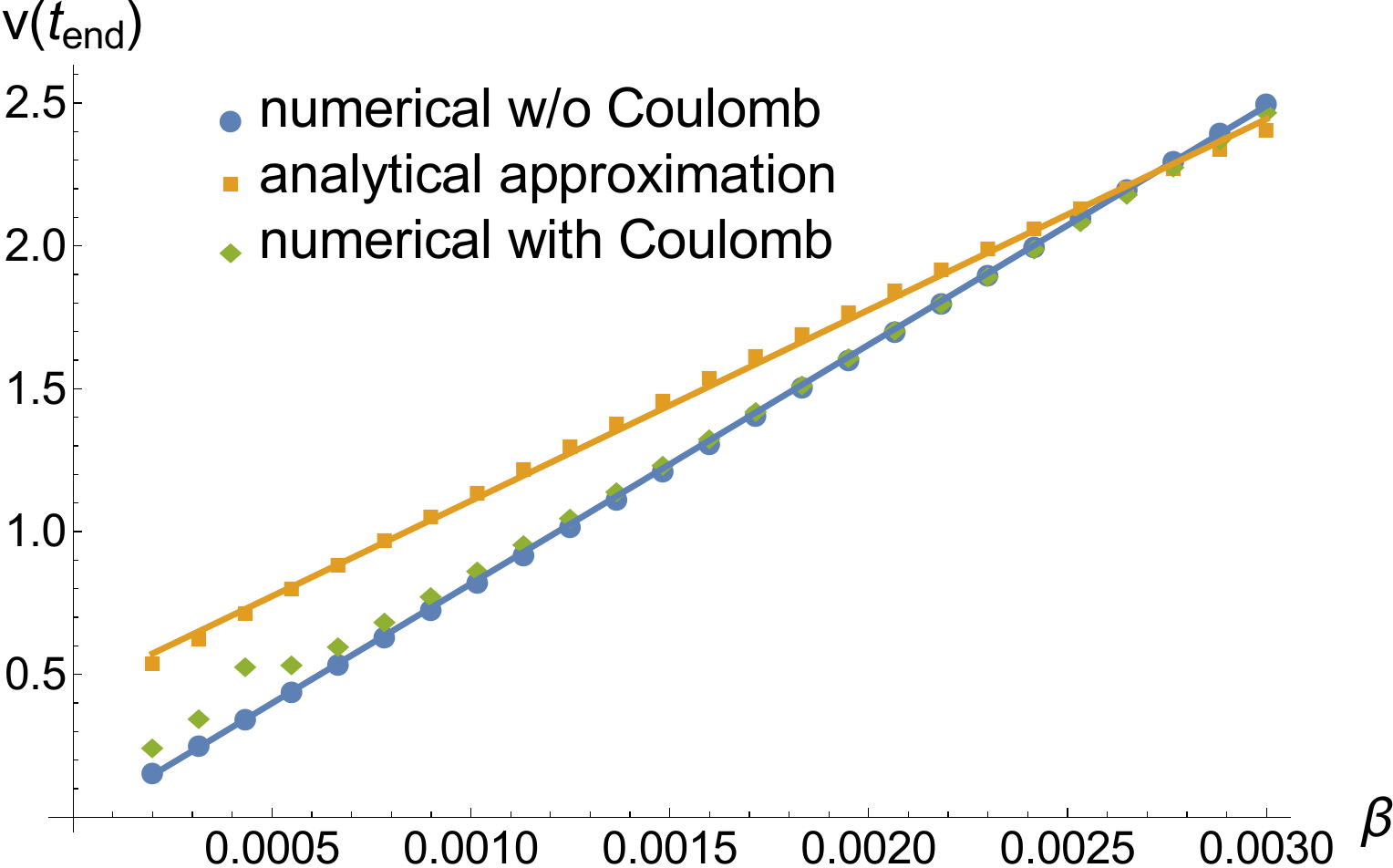}}  
\resizebox{3in}{!}{\includegraphics[angle=0]{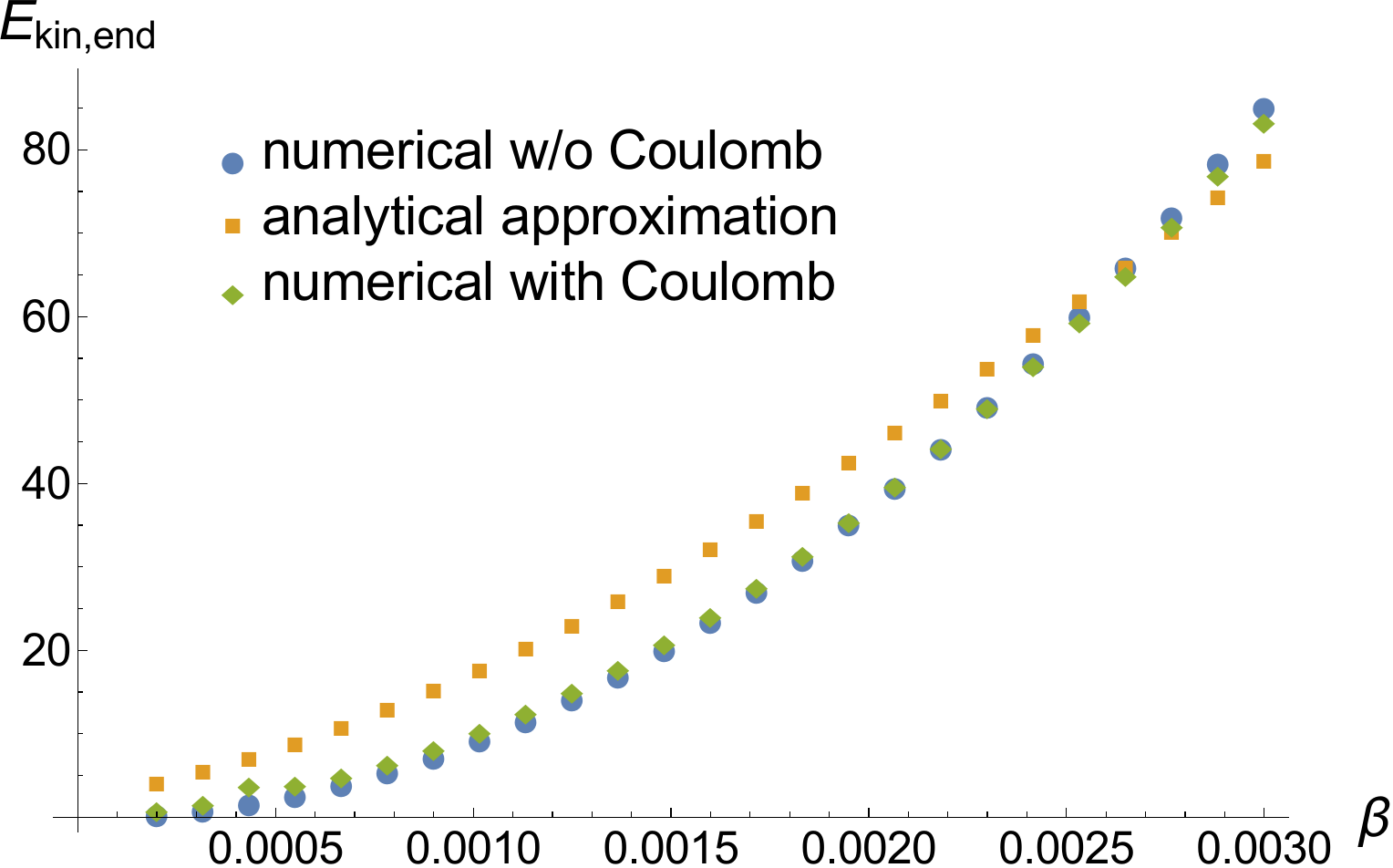}}  
\caption{The final electron velocity (top panel) and the final kinetic energy (bottom panel) as a function of the inhomogeneity parameter $\beta$ for electrons ionized at the center of the pulse ($t_0=0$). In both cases the parameters given in the caption of Fig. \ref{fig:acceleration_comparison} are used. 
The blue dots show the result obtained by solving eq. \ref{eq:ODE1} numerically and for the green diamonds the Coulomb force was added to the right hand side of this equation. The analytical solution according to eq. \ref{eq:v_end} is represented by orange squares.}
\label{fig:velocity_energy_end_vs_beta}
\end{figure}

It turns out that estimating the final velocity and thus also the kinetic energy by the analytical approximation is in good agreement with the result we obtain from the numerical solution of eq. \ref{eq:ODE1} not only for the example presented above but for a wide range of inhomogeneity parameters as Fig \ref{fig:velocity_energy_end_vs_beta} shows. In the two plots shown, the electron is ionized at $t_0=0$, corresponding to the center of the pulse. Note that this is the instance of most likely ionization \cite{ammosov1986tunnel,ppt} and that these electrons accumulate at the maximum of the HES peak.
Furthermore, the results in Fig. \ref{fig:velocity_energy_end_vs_beta} justify the initial drop of the Coulomb term in the propagation since the solution of eq. \ref{eq:ODE1} (blue dots) and the equation of motion including the Coulomb potential (eq. \ref{eq:ODE0}, green dots) differ by maximally  2.2 eV.

However, even though the analytical approximation of the velocity lead to a closed form expression, these terms are too complicated to directly derive simple dependencies. Nonetheless, the analytical derivation and the physical insights we gained from it, will help find a simple scaling law for the final energy at which the electrons accumulate, as we will see in this section, and a scheme to quickly estimate which inhomogeneity minimizes the peak width, which we will focus on in the following section.

From the data depicted in the top panel of Fig. \ref{fig:velocity_energy_end_vs_beta}, we can already guess that the scaling of the final velocity as a function of field inhomogeneity, $\beta$, has a linear dependence.  To investigate this further, we perform a power law fit $A \beta^B$ with fitting parameters A and B on the numerical solution of eq. \ref{eq:ODE1} for the final electron velocity.  Note that this power law is consistent with the fact that for $\beta =0$, the case of a homogeneous field, we expect that an electron ionized at the centre of the pulse will end up with a negligibly small velocity. 
We find that such a power law fit gives $B=1.01$ at an intensity of $I=1 \cdot 10^{14}~\mathrm{W/cm^2}$ for a range of $\beta$ between $0$ and $0.003$, which substantiates the assumption of a linear dependence of the final velocity on the inhomogeneity parameter. The insights gained from the analytical approximation can now help us understand how this linear scaling comes about.

\begin{figure}
\resizebox{3in}{!}{\includegraphics[angle=0]{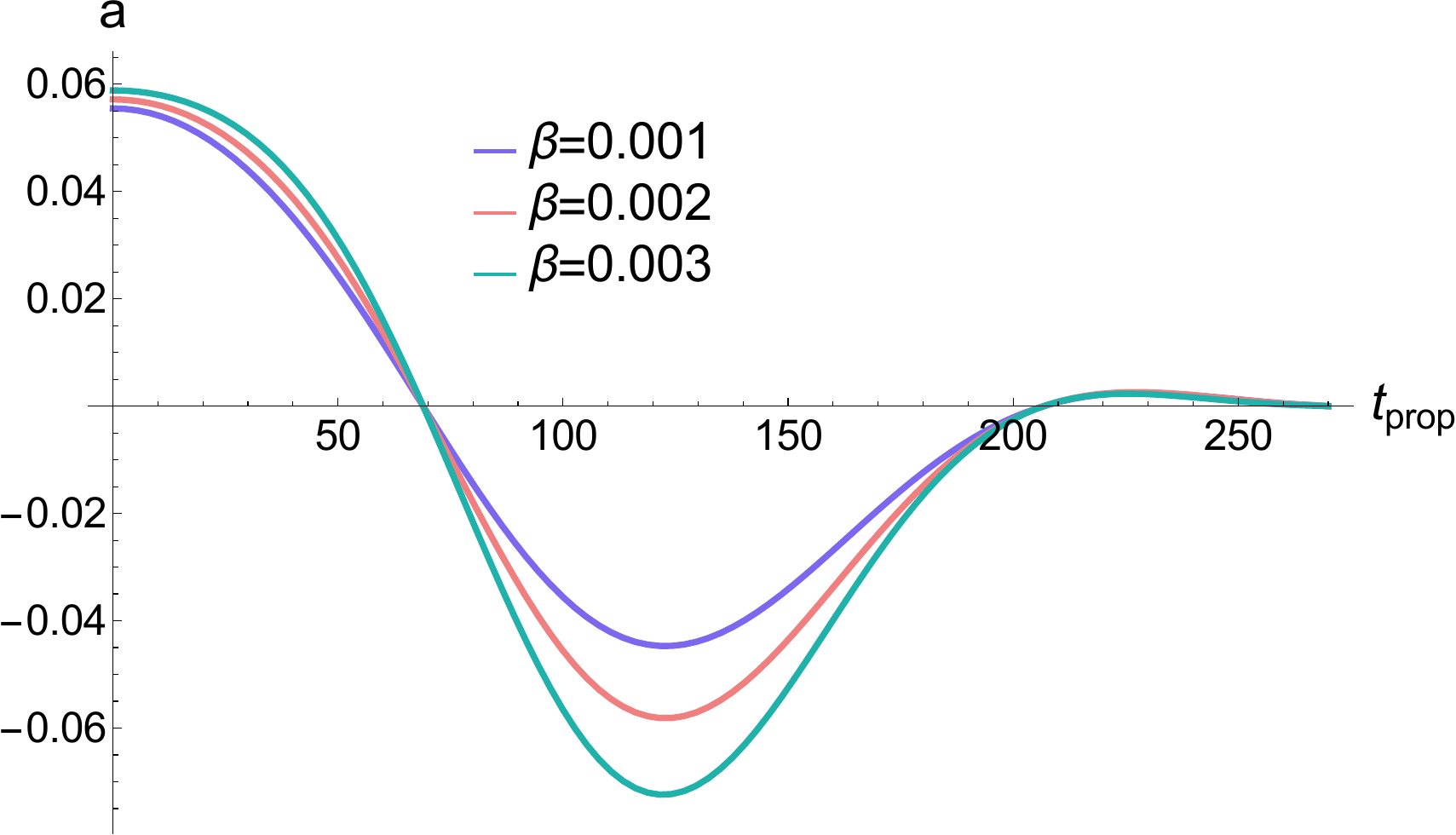}}  
\resizebox{3in}{!}{\includegraphics[angle=0]{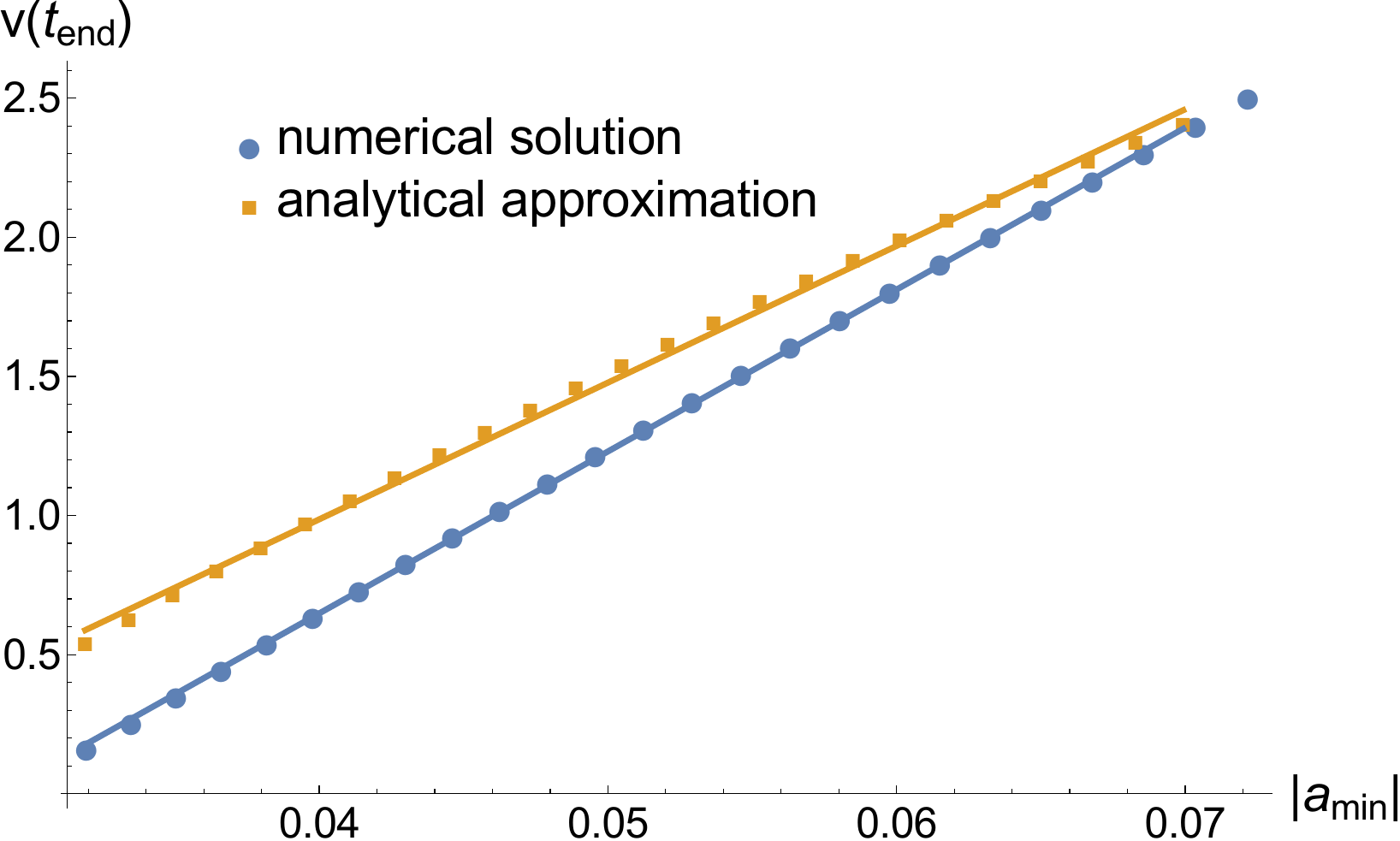}} 
\caption{Top panel: The numerical solution for $a(t)$ according to eq. \ref{eq:ODE1} for three different beta values where all the other parameters were chosen as in Fig. \ref{fig:acceleration_comparison}. The area under this curve, which equals the final electron's velocity, is influenced by $\beta$ mainly via the amplitude of the oscillations in $a(t)$. Bottom panel: The final electron velocity as a function of the modulus of the amplitude $a_{min}$ of $a(t)$ according to the numerical solution of eq. \ref{eq:ODE1} (blue dots) and according to the analytical approximation (orange dots). The straight lines are linear fits which confirm that the velocity depends linearly on $a_{min}$.}
\label{fig:aNum_vs_tProp_for_3_betaValues}
\end{figure}

To this end, we first take a look at how the acceleration $a(t)$ depends on $\beta$. After all, the velocity is nothing but the area under this curve. As the top panel in Fig. \ref{fig:aNum_vs_tProp_for_3_betaValues} shows, the roots  of $a(t)$ are scarcely affected by $\beta$ and the inhomogeneity mainly influences the amplitude of the oscillation. As the curve between the two roots bears a strong resemblance to a simple sine or cosine function, the area under this part of the curve approximately scales linearly with the amplitude. (For an intuitive understanding of this concept one may think of the integral over a half-cycle sine function with amplitude $C$, $\int_{0}^{\pi} C \sin( x) dx = C \cdot 2 $, which obviously depends linearly on the amplitude $C$.) 

The lower panel in Fig. \ref{fig:aNum_vs_tProp_for_3_betaValues} shows that indeed the final velocity scales linearly with the minimum of the acceleration $a_{min}$, which determines the area under the second hump: $v(t_{end}) \propto a_{min}$. Now, $a_{min}$ is very close to $a_1$, the acceleration taken at the point $t_1 = \pi/\omega$, where the approximation $v_{approx,1}$ began to deviate considerably from the numerical solution of eq. \ref{eq:ODE1}. 

Looking at eq. \ref{eq:ODE2}, it is clear that the acceleration is proportional to the inhomogeneity $\beta$, while $x_M(t_1)$ is approximately independent of $\beta$. Thus, we directly attain the relation $a_{min} \propto \beta$ and using the relation $v(t_{end}) \propto a_{min}$ we can infer the approximate proportionality $a_{min} \propto \beta \Rightarrow v(t_{end}) \propto \beta$ or 
\begin{equation}
  E_{kin}(t_{end}) \propto \beta^2,
\end{equation}
which confirms our finding from the purely numerical calculations.

\begin{figure*}
\resizebox{7in}{!}{\includegraphics[angle=0,width=1.0\textwidth]{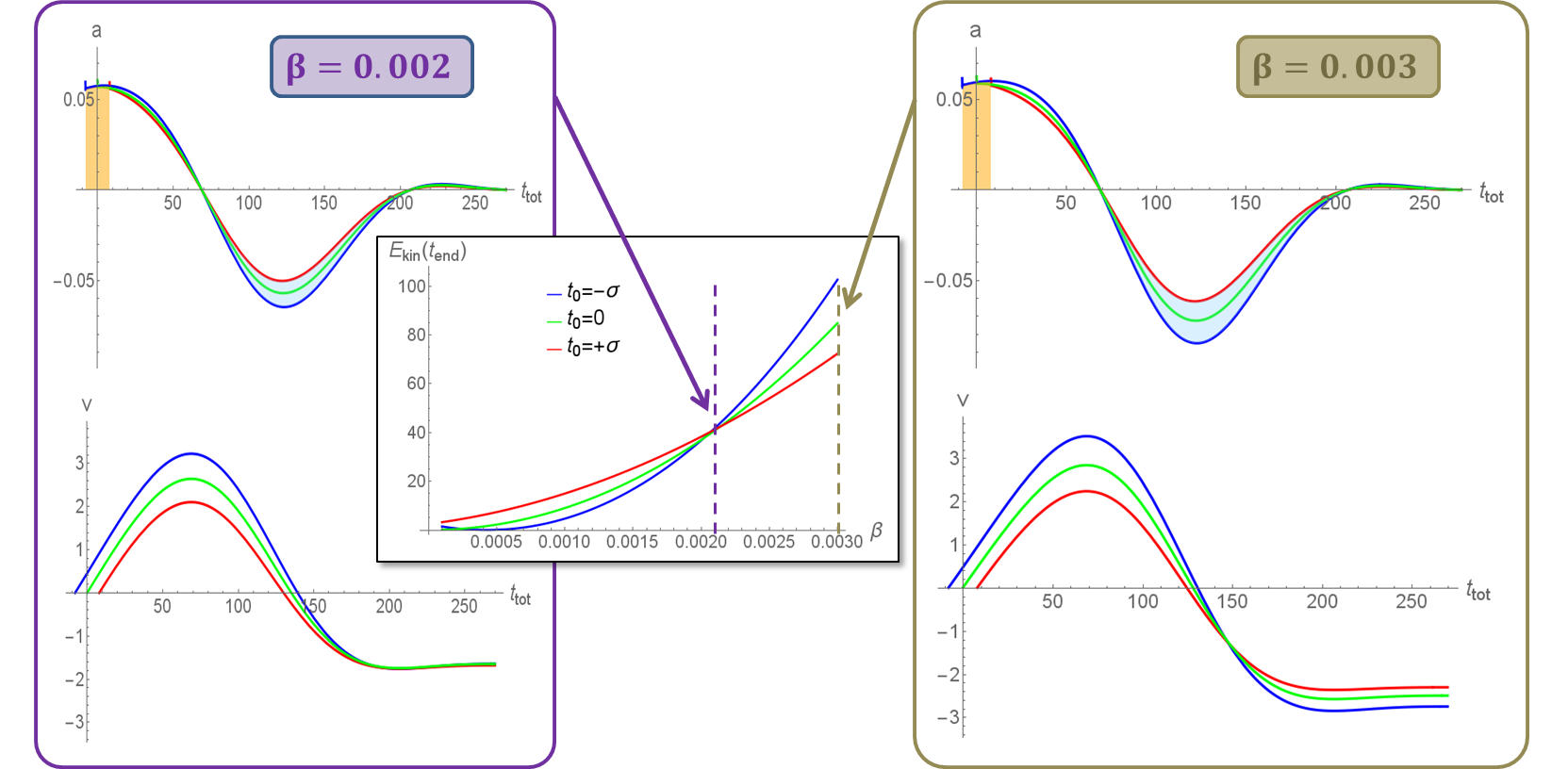}}  
\caption{The final kinetic energy (central panel), the acceleration (upper panels) and the velocity (lower panels) as a function of time according to the numerical solution of eq. \ref{eq:ODE1}. The electrons are ionized at $t_0=0$ (green lines), $t_0=-\sigma$ (blue lines), and $t_0=+\sigma$ (red lines).
Both panels on the left show results for $\beta=0.002$, whereas on the right hand side $\beta=0.003$ was used. In the central panel the final energy is shown over a large range of inhomogeneities. 
The HES peak width is assumed to be minimal at the inhomogeneity for which ionization before or after the center of the pulse has the least effect on the final kinetic energy, which is the case for $\beta \approx 0.002$ here.}
\label{fig:HES_peak_width_beta_sigma_overview}
\end{figure*}

\section{Dependence on the HES peak width on the inhomogeneity parameter} \label{sec:beta_dependence_of_HES_peak_width}
As Fig. \ref{fig:HES_peak_for_different_beta}  already revealed in a qualitative fashion, not only the center of the HES but also its width strongly depends on the inhomogeneity parameter $\beta$. Reducing the HES width is potentially of great importance to its applications in the creation of near monoenergetic electron beams. For the determination of the energy at which the HES peak is centered in section \ref{sec:beta_dependence_of_final_energy}, we had a closer look at trajectories of electrons ionized at the center of the pulse $t_0=0$ as here the ionization probability is largest. The distribution of ionization times in the adiabatic approximation reads
\begin{equation}
 P(t_0) \propto \exp \left( - \frac{2 (2I_p(t_0))^{3/2}}{|3 F(t_0)|}  \right)
\end{equation}
and using the approximations $I_p(t_0) \approx I_p(0)$ and $F(t_0) \approx E_0 \cos( \omega t_0)$ when expanding the time-dependent exponent for $\omega t \ll 1$ into a Taylor series up to second order will lead to a Gaussian distribution with standard deviation $\sigma = -\sqrt{\frac{3 E_0}{2^{5/2} I_p^{3/2}}}/\omega$ \cite{PhDThesisCornelia, Cornelia_momenta, popov2005imaginary}. Now, as we are interested in the width of the HES peak, we analyze electrons that are ionized at $t_0 = -\sigma$ and $t_0=+\sigma$ in order to be able to quantify how deviating from $t_0=0$ changes the final energy. Obviously, we want a deviation from $t_0=0$ to have the least effect on the final energy in order to minimize the HES peak width. 

The central plot in Fig. \ref{fig:HES_peak_width_beta_sigma_overview} shows the final energy for the three mentioned ionization times over a range of inhomogeneity parameters. Here, the peak width is optimal for $\beta\approx 0.002$. Note that the fact all three curves intersect in one point does not mean that the peak can be assumed to be infinitely sharp. As Fig. \ref{fig:HES_peak_for_different_beta} shows for the case of $\beta=0.002$, even though the peak is narrow it has finite width. Tracing electrons ionized at $t_0=\pm \sigma$ is only a good measure to estimate how the widths at different inhomogeneities compare to each other, but it does not allow for an estimation of the total width. 

\begin{figure}
\resizebox{3in}{!}{\includegraphics[angle=0]{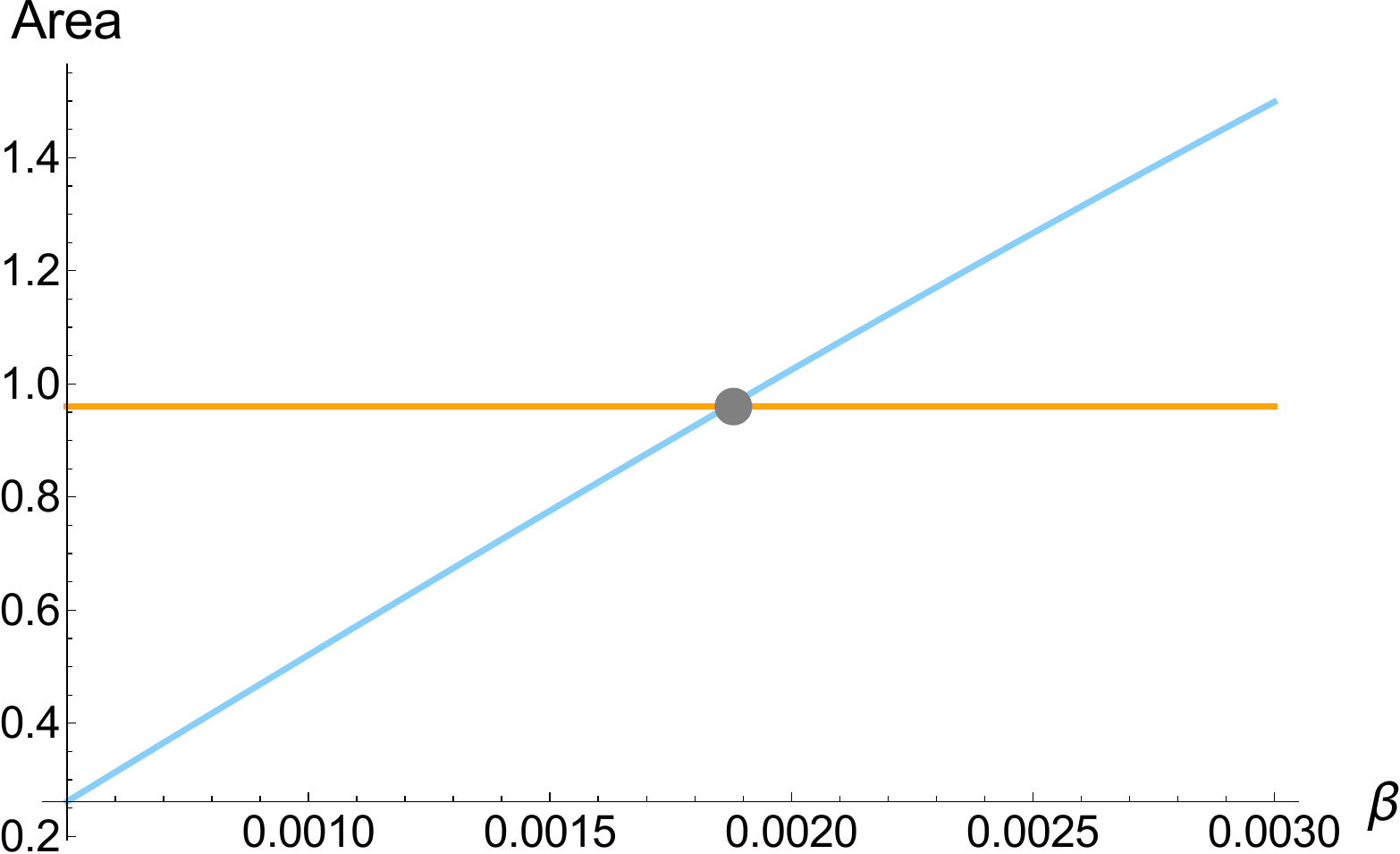}}   
\caption{Scheme how to estimate graphically the optimal $\beta$ value, which is found at the intersection of the orange line representing the `head start area' and the blue curve which represents the area between the negative second humps of the trajectories launched at $t_0=-\sigma$ and $t_0=+\sigma$. Both curves intersect at $\beta \approx 0.0019$, which serves as a good estimate for the inhomogeneity at which the minimal HES peak width is obtained.}  
\label{fig:Compensating_areas_vs_beta_Aref_for_beta=0,003}
\end{figure}

In order to understand the result observed in the central plot of Fig. \ref{fig:HES_peak_width_beta_sigma_overview}  and in order to be able to optimize the HES width at other intensities, frequencies or atoms in a quick fashion without solving the full ODE over a range of inhomogeneities, we look at the acceleration and velocity during propagation for electrons ionized at $t_0=0$ and $t_0 = \pm \sigma$. For $\beta = 0.002$, these quantities are depicted in the left panel of Fig. \ref{fig:HES_peak_width_beta_sigma_overview}. Note that here the absolute time $t_{tot}$ and not the time after ionization, $t_{prop}$, is displayed on the x-axis. Now, it is easy to see that the head start for the electron ionized at $t_0=-\sigma$ will add some positive 'area' (marked in orange) under the acceleration curve as compared to the trajectory launched at $t_0=+\sigma$. However, as  this is the very inhomogeneity that optimizes the pulse width, $\beta=0.002$, this additional positive area under the acceleration curve is compensated by the additional negative area (light blue shaded area) under the second hump. Thus, in this case, the final total area under the curve and hence the final velocity is about the same for electrons ionized at $t_0=0, -\sigma$ as well as $+\sigma$. This is not the case for e.g. $\beta = 0.003$, which is shown in the right panel of Fig. \ref{fig:HES_peak_width_beta_sigma_overview}. Here, the positive area due to the head start of the electron launched at $t_0=-\sigma$ as compared to the one launched at $t_0=+\sigma$, shaded in orange, is of approximately the same size as it was the case for $\beta=0.002$. In contrast, the difference in the negative area under the second hump of the acceleration curve (shaded in blue) has grown with $\beta$ and the overall area under the accerlation curves for $t_0=-\sigma$ is more negative than it is for $t_0=+\sigma$. Thus, the latter ends up at a smaller kinetic energy. The same concept can be used for arbitrary inhomogeneities: The orange `head start area' $A_{headstart}$ can be approximated to be independent of $\beta$ and can be estimated by assuming a rectangle of a width in time of $2 \sigma$. The rectangle's height can be estimated easily by plugging in the initial conditions for $t_0 = 0$ into eq. \ref{eq:ODE1}. This rectangle's area is represented as an orange line in Fig. \ref{fig:Compensating_areas_vs_beta_Aref_for_beta=0,003}.  The negative area, $A_{neg}$, under the second hump is shown by a blue curve in Fig. 8 
(for derivation, see Appendix \ref{sec:appendix_beta_minimal_HES_width_scheme}). The minimum width of HES occurs where the orange and the blue curves intersect, near $\beta \approx 0.002$, signifying the convergence in energy of trajectories starting at different ionization times.

\section{Conclusion} \label{sec:Conclusions}
The numerical results for the electron's motion in the superposed potential of an inhomogeneous laser field, as it typically occurs close to a nano-structure, and the Coulomb field of the ionized atom are found to be dominated by the inhomogeneous field and agree well with the results where the Coulomb potential was neglected.  In particular, we found that neglecting the Coulomb field and assuming a slowly-varying pulse envelope (relative to the frequency of the laser field), an analytic approximation for the acceleration can be found using Mathieu functions.  For the accurate description of final electron energy for sufficiently short pulses another approach was needed due to the break-down of the slow-varying envelope assumption away from the center of the pulse. 
However, splitting the time of propagation into a first part, where the time-dependence of the envelope was neglected, and into a second part, where only the envelope was taken into account (neglecting the fast oscillations), good agreement with the full numerical solution for electron velocity and hence final kinetic energy was obtained. Most importantly, the knowledge about the electron's movement in the inhomogeneous field obtained from the analytical studies allowed for a derivation of a simple scaling law stating that the spectral position of the HES maximum depends quadratically on the inhomogeneity parameter $\beta$. Moreover, it was found that there is an inhomogeneity at which the width of the HES peak is minimal and a scheme to estimate this optimal inhomogeneity based on the analytical descriptions of the velocity was presented.

\begin{acknowledgments}
We thank Cornelia Hofmann for valuable discussions. L.O. acknowledges support by the Max Planck Society via the IMPRS 'Dynamical Processes in Atoms, Molecules and Solids' and A.S.L. is supported by the Max Planck Center for Attosecond Science (MPC-AS).
\end{acknowledgments}

\appendix 
\section{Partial integration for analytical approximation of the velocity} \label{sec:appendix_integration_by_parts}
We approximate the velocity by integrating the acceleration we obtained from plugging in the solution given in eq. \ref{eq:x_Mathieu} into the RHS of eq. \ref{eq:ODE2}. As this expression for the acceleration equals $a_M(t) \cdot f(t)$, where  $a_M$ denotes the second derivate of $x_M$ (eq. \ref{eq:x_Mathieu}) with respect to time, we can integrate by parts in the following way: 
\begin{align}
\begin{split}
 &v_{approx1}(t) = \int_{t_0}^{t} a_M(t') \cdot f(t') dt' \\
 &= \left[ v_M(t') \cdot f(t') \right]_{t_0}^{t} - \int_{t_0}^{t} v_M(t') \cdot \frac{d}{dt'} f(t') dt'\\
 &\approx \left[ v_M(t') \cdot f(t') \right]_{t_0}^{t} - \int_{t_0}^{t} (v_M \cdot f(t')) \cdot \frac{d}{dt'} f(t') dt'\\
 &=  \left[ v_M(t') \cdot f(t') \right]_{t_0}^{t}  - \left( \left[ x_M(t') \cdot  f(t') \cdot \frac{d}{dt'} f(t') \right]_{t_0}^{t} - V  \right), \label{eq:vApprox1_appendix}
 \end{split}
\end{align}
where, for further improvement of the approximation, in the third line $v_M$ was replaced by $v_M \cdot f$ as the latter is the lowest order approximation (first summand in lines 2 to 4) of $v_{approx1}$. Moreover, $V$ denotes $\int  x_{M}(t') \left(  f(t') \cdot \frac{d^2}{dt'^2} f(t') + \left( \frac{d}{dt'} f(t') \right)^2\right) dt'$, which we drop assuming a slowly-varying envelope.

\section{Scheme to estimate $\beta$ for minimal HES width} \label{sec:appendix_beta_minimal_HES_width_scheme}
In order to estimate the negative area $A_{neg}$ under the second hump we do a trick similar to the one done to estimate the $\beta$ dependency of the energy at which the HES peak is centered (see section \ref{sec:beta_dependence_of_final_energy}): As the roots of the acceleration are not affected by $\beta$ and as the function looks close to something like a sine or cosine, we can assume that the area scales linearly with the minimum $a_{min}$ of the curve: $A_{neg} \propto |(a_{t0=-\sigma}(t_{min}) - a_{t0=+\sigma}(t_{min}))|$. However, as we need the absolute number of $A_{neg}$, we also need the prefactor $A_{ref}$ which determines the slope: 
\begin{equation}
 A_{neg} = A_{ref} |a_{min,t0=-\sigma}(t_{min}) - a_{min,t0}=+\sigma(t_{min})|.
\end{equation}
As this equation calculates just the difference between the full area under the second hump for $t_0=-\sigma$ and the full area under the second hump for $t_0=+\sigma$, this prefactor can be determined by the knowledge about the area under the second hump for a single ionization time. We choose $t_0=0$ as for this case we can use eq. \ref{eq:vApprox1} and eq. \ref{eq:v_end} to determine the area under the second hump immediately as $|v_{end}(t_{z,1}) - v_{Approx,1}(t_{z,2})|$ where $t_{z,1/2}$ denote the limits in time of the negative hump, so the roots of $a(t)$. Thus, we get
\begin{equation}
 A_{ref} \approx  \frac{|v_{end}(t_{z,1}) - v_{Approx,1}(t_{z,2})|}{|a_{t_0=0}(t_{min})|} 
\end{equation}

In Fig. \ref{fig:Compensating_areas_vs_beta_Aref_for_beta=0,003} one can find $A_{neg}$ as a function of $\beta$. Now, the inhomogeneity at which the head start area (orange line) and the negative Area $A_{neg}$ (blue line) intersect, should be a good estimate for the inhomogeneity to choose for a minimal width in energy of the HES peak. In the case shown in Fig. \ref{fig:Compensating_areas_vs_beta_Aref_for_beta=0,003}, this graphical method gives $\beta\approx 0.0019$ for the optimal inhomogeneity, which agrees well with $\beta \approx 0.002$, which we got from solving. eq. \ref{eq:ODE1} numerically for $t_0=0,\pm \sigma$ for various $\beta$ (see central panel in Fig. \ref{fig:HES_peak_width_beta_sigma_overview}). 



\end{document}